\documentclass[aps,prb,twocolumn,superscriptaddress,longbibliography]{revtex4-2}

\usepackage{amsmath,amssymb}
\usepackage[pdftex]{hyperref,graphicx}
\hypersetup{colorlinks = true, urlcolor = blue, linkcolor = blue, citecolor = blue}

\usepackage{physics}
\usepackage{xcolor}
\usepackage{bm}

\newcommand{\angstrom}{\text{\normalfont\AA}}
 
\newcommand{\nn}{\nonumber \\} 
\newcommand{\tp}{ ^{\intercal} }

\begin{document}
\title{Quantized and unquantized zero-bias tunneling conductance peaks in Majorana nanowires: Conductance below and above $ 2e^2/h $}
\author{Haining Pan}
\thanks{These authors contributed equally to this work.}
\affiliation{Condensed Matter Theory Center and Joint Quantum Institute, Department of Physics, University of Maryland, College Park, Maryland 20742, USA}
\author{Chun-Xiao Liu}
\thanks{These authors contributed equally to this work.}
\affiliation{Qutech and Kavli Institute of Nanoscience, Delft University of Technology, Delft 2600 GA, The Netherlands}
\author{Michael Wimmer}
\affiliation{Qutech and Kavli Institute of Nanoscience, Delft University of Technology, Delft 2600 GA, The Netherlands}
\author{Sankar Das Sarma}
\affiliation{Condensed Matter Theory Center and Joint Quantum Institute, Department of Physics, University of Maryland, College Park, Maryland 20742, USA}

\begin{abstract}
Majorana zero modes can appear at the wire ends of a 1D topological superconductor and manifest themselves as a quantized zero-bias conductance peak in the tunneling spectroscopy of normal-superconductor junctions.
However, in superconductor-semiconductor hybrid nanowires, zero-bias conductance peaks may arise owing to topologically trivial mechanisms as well, mimicking the Majorana-induced topological peak in many aspects.
In this work, we systematically investigate the characteristics of zero-bias conductance peaks for topological Majorana bound states, trivial quasi-Majorana bound states and low-energy Andreev bound states arising from smooth potential variations, and disorder-induced subgap bound states. 
Our focus is on the conductance peak value (i.e., equal to, greater than, or less than $2e^2/h$), as well as the robustness (plateau- or spike-like) against the tuning parameters (e.g., the magnetic field and tunneling gate voltage) for zero-bias peaks arising from the different mechanisms.
We find that for Majoranas and quasi-Majoranas, the zero-bias peak values are no more than $2e^2/h$, and a quantized conductance plateau forms generically as a function of parameters. By contrast, for conductance peaks due to low-energy Andreev bound states or disorder-induced bound states, the peak values may exceed $2e^2/h$, and a conductance plateau is rarely observed unless through careful {postselection and fine-tuning}.
Our findings should shed light on the interpretation of experimental measurements on the tunneling spectroscopy of normal-superconductor junctions of hybrid Majorana nanowires.
\end{abstract}
\maketitle
\section{Introduction}
Majorana zero modes (MZMs), which are the fundamental non-Abelian units for topological quantum computing,~\cite{kitaev2001unpaired,kitaev2003faulttolerant,freedman2003topological,dassarma2005topologically,nayak2008nonabelian} have been extensively studied in experiments on superconductor-semiconductor (SC-SM) hybrid nanowires over the past decade~\cite{mourik2012signatures,das2012zerobias,deng2012anomalous,churchill2013superconductornanowire,finck2013anomalous, deng2016majorana,nichele2017scaling,zhang2017ballistic,gul2018ballistic,kammhuber2017conductance,vaitiekenas2018effective,moor2018electric,zhang2021large,*zhang2018quantizeda,bommer2019spinorbit,grivnin2019concomitant,anselmetti2019endtoend,menard2020conductancematrix,puglia2020closing,pan2020situ}.
Owing to the improved sample quality of hybrid nanowires, several hallmarks of Majorana zero modes have been observed in tunneling conductance spectroscopy measurements, even the apparent quantized zero-bias conductance peak (ZBCP). Recently, Refs.~\onlinecite{nichele2017scaling, zhang2021large,zhang2018quantizeda} have shown an almost-quantized conductance of $ 2e^2/h $ with plateaus robust against the Zeeman field or the gate voltage to some degree. However, these experiments are inconclusive to confirm the MZMs due to the lack of the manifestation of other hallmarks of MZMs, which should be accompanied by the observation of the quantized ZBCP at the topological quantum phase transition (TQPT), such as the bulk gap closing and reopening features~\cite{stanescu2012close,huang2018metamorphosis} and the growing Majorana oscillations with increasing magnetic field strength~\cite{dassarma2012splitting}. Also, no nonlocal correlation measurements have been reported as expected for a topological state. Therefore, it is necessary to understand the origin of the plateau in the conductance data before making any conclusions.

MZMs can induce a quantized conductance of $ 2e^2/h $~\cite{sengupta2001midgap,law2009majorana,flensberg2010tunneling,wimmer2011quantum} in a long wire due to perfect Andreev reflection in the local conductance measurement in the {tunneling} spectroscopy experiment. However, this `quantized conductance of $ 2e^2/h $' is only a necessary condition but not a sufficient one to deduce the topological MZM. Many other mechanisms can also trivially induce seemingly quantized ZBCPs arising from subgap Andreev bound states (ABSs)~\cite{kells2012nearzeroenergy,prada2012transport,liu2017andreev,liu2018distinguishing,moore2018twoterminal,penaranda2018quantifying,reeg2018zeroenergy,vuik2019reproducing,pan2020physical,moore2018quantized}. For example, inhomogeneous chemical potential and random disorder are common mechanisms that may induce the trivial ABS with $ \sim 2e^2/h $ ZBCPs~\cite{pan2020physical}. However, unlike the robust topological MZM-induced quantized ZBCPs, these trivial ZBCPs are usually either unstable or unquantized. An important issue in this context is the extent to which these trivial ZBCPs could accidentally produce, perhaps through some fine-tuning of parameters, somewhat stable-looking $ 2e^2/h $ apparent quantization, consequently misleading the experimentalists into thinking that topological MZMs might have been observed.  The crucial question is whether one can assert that the observation of a fine-tuned apparently stable quantized ZBCP automatically implies the existence of topological MZMs.

The answer to this question turns out to be negative. The inhomogeneous smooth confining potential is a typical counterexample that can manifest trivial stable quantized ZBCPs with a nonzero probability given appropriate parameters. 
{Namely, although the hybrid nanowire is topologically trivial deep inside its bulk, the barrier-like confining potential near the wire end serves as a domain wall for the two spin-polarized channels and induces a Majorana at the domain wall for each channel. 
Owing to the smoothness of the inhomogeneous potential, despite the overlapping wave functions, the two Majoranas do not couple with each other effectively, thus forming a near-zero-energy fermionic quasi-Majorana zero mode~\cite{vuik2019reproducing, moore2018twoterminal,liu2019conductance}.}
The nonlocality of Majoranas is not preserved in such a situation; the two Majoranas can both couple to the same normal lead at the end of the nanowire. Therefore, it manifests a robust ZBCP with the conductance slightly below the theoretical quantized value of $ 2e^2/h $. Even if quasi-Majoranas are not the topological MZM, if the experimentalists observe such robust ZBCP with the conductance slightly below $ 2e^2/h $, it is still a positive development because it may indicate the real MZMs are not far away---one only has to reduce the inhomogeneity along the nanowire to separate the two overlapping MZMs in order to see the topological MZMs. In addition, finite temperature and dissipation in experiments cause an inevitable broadening, which lowers the conductance peak. Therefore, it is conceivable that the expected topological ZBCP has a robust conductance peak that is slightly below $ 2e^2/h $ because of broadening, even if it is purely induced by the real MZM. In this sense, it may not be too worrisome to observe a robust plateau with the conductance slightly below $ 2e^2/h $ in experiments. Such a peak may arise from topological MZMs or quasi-MZMs, and the quasi-MZMs should generate topological MZMs if the two overlapping MZMs can be spatially separated.

However, what if the observed conductance is slightly above $ 2e^2/h $? Is it possible that an inhomogeneous-potential-induced robust ZBCP with conductance slightly above $ 2e^2/h $ could arise from trivial quasi-MZMs? In this work, we show that it is, in principle, possible to find trivial ZBCPs induced by an inhomogeneous potential that have conductances above $ 2e^2/h $, but they are all in the form of spikes and not robust as a function of parameters. We discuss two such explicit examples of the inhomogeneous potential and find that they are both dip-like inhomogeneous potentials (i.e., a potential well rather than a potential barrier, which we call a `dip' throughout) in contrast to the barrier-like smooth confining potential which can induce a robust plateau that may have a conductance slightly below $ 2e^2/h $. Therefore, this finding leads to the fact that, if the observed ZBCP manifests a robust conductance plateau that is slightly above $ 2e^2/h $, it is very unlikely to be induced by the inhomogeneous potential, irrespective of the shape--- barrier or dip--- because, otherwise, its conductance should be below $ 2e^2/h $ or in the form of spikes instead of a robust plateau. Such a `spiky' ZBCP rises quickly to $ 4e^2/h $ and then falls rapidly to almost zero without manifesting any stability as a function of parameters in contrast to the robust $ 2e^2/h $ (or slightly below it) ZBCP often induced by the barrier-type potential inhomogeneity.  We establish that a ZBCP plateau slightly above $ 2e^2/h $ cannot, therefore, arise from quasi-MZMs or topological MZMs which would produce ZBCPs at or slightly below $ 2e^2/h $.

Finally, we show that even if such a robust plateau with a conductance above $ 2e^2/h $ cannot be induced by the inhomogeneous potential, it does not preclude the possibility of a robust plateau with a conductance above $ 2e^2/h $ arising from other mechanisms. We show that such a scenario may occur in the presence of on-site random disorder in the chemical potential through { postselections} of the random configuration. 
{Here, { postselections} means that only about 8\% of the disorder realizations in our simulation give rise to a robust conductance plateau above $ 2e^2/h ${ , and it is only after carrying out the simulations we know which disorder configurations produce the desired outcome of $ 2e^2/h $  peaks--- there is no way to know the outcome beforehand.}}
However, even if we allow careful fine-tuning of the random configuration, the robustness is still delicate. Unlike the inhomogeneous smooth confining potential, which can manifest a robust conductance plateau (but not above $ 2e^2/h $) in a large region of the parameter space, the robust region induced by the random disorder is rather limited: It is robust only against a small range of energy interval of the Zeeman field, and is not robust against all the gate voltages, e.g., it may manifest robustness against the Zeeman field to some degree, but it becomes uncontrollable when the tunneling gate voltage is tuned. Therefore, if such a plateau with the conductance above $ 2e^2/h $ occurs in the laboratory, it indicates that the observed ZBCP cannot be the bad ZBCP induced by inhomogeneous potential, as introduced in Ref.~\onlinecite{pan2020physical}; instead, it must be the ugly ZBCP arising from disorder, and such a ZBCP with a value above $ 2e^2/h $ is unlikely to be very robust.  It is significant in this context to point out that the recent experiments~\cite{nichele2017scaling,zhang2021large,zhang2018quantizeda} reporting $ 2e^2/h $ conductance quantization invariably find a conductance slightly above $ 2e^2/h $ indicating that the current experimental nanowires are disorder-limited.

{ It may be useful to point out that all experimental reports of the observation of zero-bias peaks, and particularly, ZBCPs with $\sim 2e^2/h$ conductance, involve extensive postselection and fine-tuning.  Only a few samples show ZBCPs in narrow ranges of experimental gate voltages, tunnel barriers, and magnetic fields, and the experimental procedure is to post-select these few samples ($ \sim 2-3\%$ success rates typically) and then fine-tune gate voltages and tunnel coupling in order to find approximate $ 2e^2/h $ , and then report those peaks in the experimental papers.  Most of the tunneling conductance measurements on most nanowires simply fail to produce anything notable, and the experimental protocol is to do postselection and fine-tuning to find ZBCPs which are publication-worthy.  Our theoretical procedure of post-selecting disorder configuration to find $ 2e^2/h $  ZBCPs is thus consistent with the current experimental protocol.  Also, experimentally, the inhomogeneous potential is unintentional and hence unknown, often arising from accidental quantum dots formed at the nanowire ends or unintentional inhomogeneous potentials arising from applied gate voltages during fine-tuning.  Thus, our using and fine-tuning model inhomogeneous potentials in order to fine-tune and post-select desirable ZBCPs is also consistent with the current experimental protocol.  It is of course entirely possible that such protocols lead to considerable confirmation bias in the experiments since one knows precisely what one is looking for, and is fine-tuning and post-selecting to find theoretically predicted ZBCPs of the correct magnitude.}

The remainder of the paper is organized as follows. In Sec.~\ref{sec:model}, we introduce the Hamiltonian of the SC-SM hybrid nanowire, and how the inhomogeneous and disorder potentials are modeled.  In Sec.~\ref{sec:smooth}, we present our main results of the robust (almost-) quantized conductance plateau in the presence of the inhomogeneous potential in 1D single-subband nanowire, and also in the more realistic 3D model using the Thomas-Fermi-Poisson method. In Sec.~\ref{sec:dip}, we present the opposite scenario, where the conductance does not manifest the robustness and becomes arbitrary in the presence of the dip-like inhomogeneous potential. In Sec.~\ref{sec:disorder}, we show the conductance spectra in the presence of random disorder. We discuss the implication of our results in Sec.~\ref{sec:discussion}. We conclude in Sec.~\ref{sec:conclusion}. For completeness, we also provide two movie files showing {the} results of our simulations while scanning system parameters~\cite{movie}.

\section{Model and method}\label{sec:model}
\begin{figure}[t]
	\centering
	\includegraphics[width=3.4in]{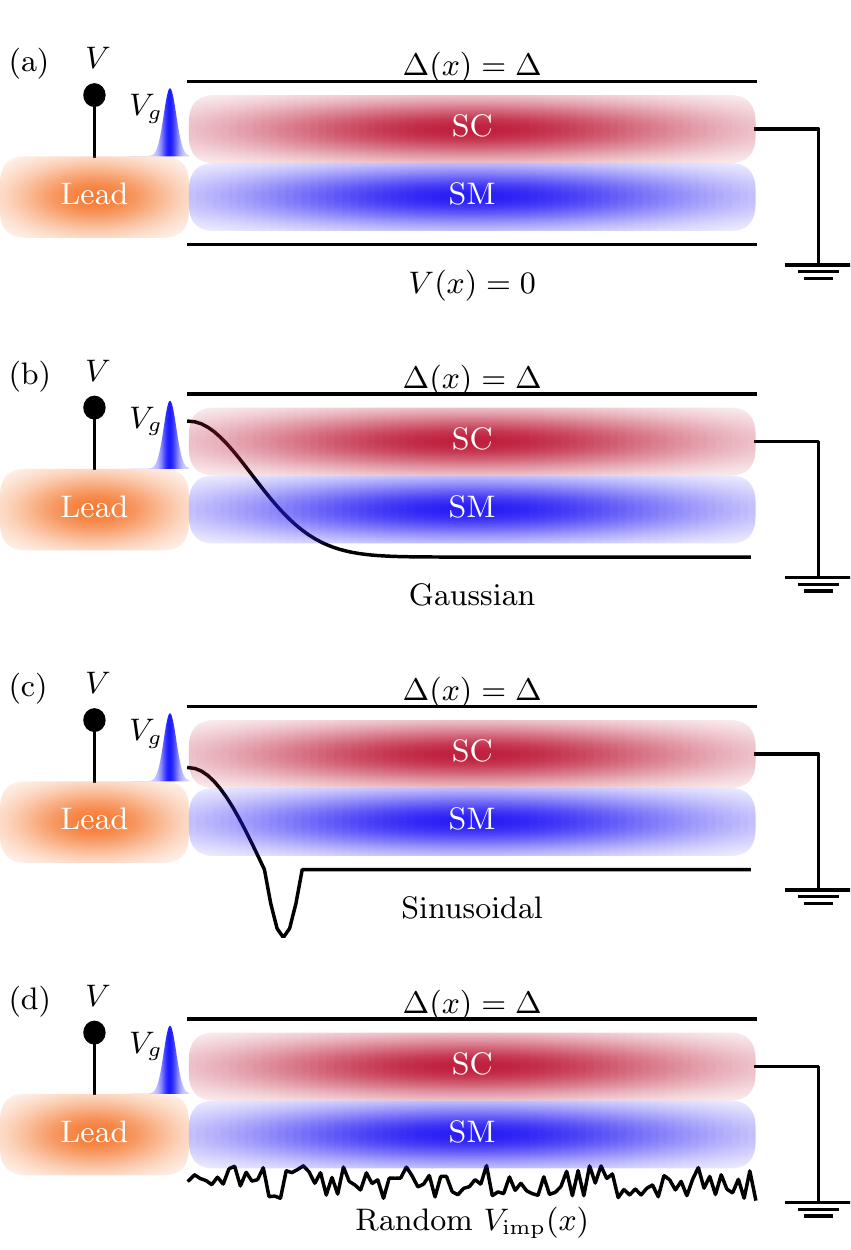}
	\caption{Schematics for the 1D NS junction model (a) of a pristine wire; (b) in the presence of Gaussian inhomogeneous potential; (c) in the presence of sinusoidal potential; (d) in the presence of random disorder. $ V_g $ is the potential barrier at the NS junction.}
	\label{fig:1}
\end{figure}

In much of this work, we model a finite-length superconductor-semiconductor hybrid nanowire using the 1D minimal model, as shown schematically in Fig.~\ref{fig:1}. The corresponding Bogoliubov-de Gennes (BdG) Hamiltonian is~\cite{lutchyn2010majorana,sau2010nonabelian,sau2010generic,oreg2010helical}

\begin{align}\label{eq:H}
&H=\frac12\int_{0}^{L} dx \hat{\Psi}^\dagger(x) H_{\text{BdG}} \hat{\Psi}(x), \nn
&H_{\text{BdG}} = H_{\text{pristine}} + H_{\text{V}}, \nn
&H_{\text{pristine}} = H_{\text{SM}}+H_{Z}+H_{\text{SC}} \nn
&= \left( -\frac{\hbar^2}{2m^*} \partial^2_x -i \alpha \partial_x \sigma_y - \mu \right) \tau_z + E_Z \sigma_x - \gamma\frac{\omega+\Delta_0\tau_x}{\sqrt{\Delta_0^2-\omega^2}}, \nn
&H_{\text{V}} = V(x)\tau_z
\end{align}
under the basis of $\hat{\Psi} = \left(\hat{\psi}_{\uparrow},\hat{\psi}_{\downarrow},\hat{\psi}_{\downarrow}^\dagger,-\hat{\psi}_{\uparrow}^\dagger\right)\tp$.
Here $H_{\text{pristine}}$ is the Hamiltonian for a pristine Majorana nanowire in which all the physical parameters are spatially homogeneous, while $H_{\text{V}}$ represents the potential inhomogeneity inside the nanowire. $L$ is the length of the hybrid nanowire. We describe each term in the Hamiltonian below.

\subsection{Pristine nanowire}

The Hamiltonian for the pristine nanowire is $H_{\text{pristine}} = H_{\text{SM}}+H_{Z}+H_{\text{SC}}$ in Eq.~\eqref{eq:H}.
$ H_{\text{SM}} $ describes the bare spin-orbit-coupled semiconducting nanowire, with $m^*$ being the effective mass of the conduction band electrons, $ \alpha $ the strength of the Rashba spin-orbit coupling, and $ \mu $ the chemical potential. $ \vec{\bm{\sigma}} $ and $ \vec{\bm{\tau}} $ are vectors of Pauli matrices acting on the spin and particle-hole spaces, respectively. 
$H_{Z}$ describes the induced Zeeman splitting due to an externally applied magnetic field parallel to the nanowire, with the field strength $E_Z$.
$H_{\text{SC}}$ describes the superconducting proximity effect induced by a conventional $s$-wave superconductor. The frequency-dependent self-energy term in $ H_{\text{SC}} $ is obtained by integrating out the degrees of freedom in the parent superconductor, with $\gamma$ being the SC-SM interfacial coupling strength producing the proximity effect, and $\Delta_0$ being the pairing potential of the parent superconductivity.
{Note that here we ignore the Zeeman-field dependence of the parent superconducting gap because such dependence does not affect the eigenenergies, eigenwavefunctions, and conductance near zero bias, and also because we want the topological ZBCP to appear at high magnetic fields.}
Near zero energy ($\omega \to 0$), the induced superconducting gap is $\gamma \tau_x$.
In the numerical calculations, unless otherwise stated, we choose the parameters corresponding to the InSb-Al hybrid nanowire:~\cite{lutchyn2018majorana} $ m^*=0.015 m_e $ ($ m_e $ is the rest electron mass), {$\mu=1$ meV,} $ \Delta_0=0.2 $ meV, $ \alpha=0.5 $ eV\angstrom, $ \gamma=0.2 $ meV, nanowire length $ L=3~\mu$m, and phenomenological tunneling barrier height $ V_g=10 $ meV, and assume the zero-temperature limit. Our generic conclusions should not depend on the choice of these parameters at all.

\subsection{Potential inhomogeneity}
The effect of potential inhomogeneity inside the hybrid nanowire is described by $ H_{\text{V}} $ in Eq.~\eqref{eq:H}.
In realistic junction devices, such an inhomogeneity may arise from the effect of gates controlling various voltages in the hybrid nanowire, from charge impurities, or unintentional disorders due to imperfect sample quality. There could also be unintentional effective quantum dots at the wire ends creating an inhomogeneous potential.
Although the precise profile of the potential inhomogeneity is not known \emph{a priori}, here, without loss of generality, we consider three different types of potential profiles, representing the typical scenarios for the potential inhomogeneity in realistic devices.

The first scenario for the potential inhomogeneity is a Gaussian-like profile near the NS junction: 
\begin{align}
V(x)=V_{\text{max}}\exp(-\frac{x^2}{2\sigma^2})
\label{eq:Gaussian}
\end{align}
with $\sigma$ being the width, and $V_{\text{max}}$ the peak of the inhomogeneity, as shown in Fig.~\ref{fig:1}(b). 
Note that the potential $V(x)$ acts as a confinement potential barrier when $V_{\max} >0$, while it becomes a potential dip when $V_{\max} <0$.
Confinement- or dip-like potential profiles have qualitatively different effects on the conductance spectroscopy of the NS junction.

In the second scenario, we consider a sinusoidal-function-like potential profile near the NS junction:
\begin{align}
&V(x)=V_{\text{max}}\cos(\frac{x\pi}{2\sigma})\theta(\sigma-x) \nn
&\quad-V_{\text{max}}^{\prime}\sin(\frac{x-\sigma}{\sigma^{\prime}}\pi)\theta(x-\sigma)\theta(\sigma+\sigma^{\prime}-x),
\label{eq:sinusoidal}
\end{align}
where $\theta(x) $ is the Heaviside step function. The sinusoidal potential contains barrier-like and dip-like parts simultaneously, as shown in Fig.~\ref{fig:1}(c).  $V_{\text{max}}$ and $\sigma$ denote the height and width of the barrier-like potential, while $ V_{\text{max}}^{\prime}$ and $ \sigma^{\prime} $ represent the depth and width of the dip-like potential.

The third scenario is a disorder-induced random potential
\begin{align}
&V(x)=V_{\text{imp}}(x), \nn
&\expval{V_{\text{imp}}(x)} = 0, \nn
&\expval{V_{\text{imp}}(x)V_{\text{imp}}(x')}=\sigma_\mu^2\delta(x-x').
\label{eq:disorder}
\end{align}
Here $V_{\text{imp}}(x)$ is a short-ranged random potential drawn from an uncorrelated Gaussian distribution with zero mean and standard deviation $\sigma_\mu$. Note that the numerical results, including the disorder-induced random potential, are based on one particular configuration of $V_{\text{imp}}(x)$, as shown in Fig.~\ref{fig:1}(d), without ensemble average. Note that while Eqs.~\eqref{eq:Gaussian} and~\eqref{eq:sinusoidal} are deterministic and smooth, the disorder potential defined by Eq.~\eqref{eq:disorder} is random and hence nondeterministic, distinguishing the smooth inhomogeneous potential from the nonsmooth random disorder potential.
{We add that in the current work, we focus on only the fluctuations in the chemical potential, and neglect such a fluctuation in other physical parameters like superconducting pairing potential, Zeeman field, and spin-orbit coupling strength, because the disorder effect in the chemical potential is the most representative one and is believed to dominate in realistic experiments.}

We mention that the disorder strength discussed in this paper should not be too large compared to the chemical potential $ \mu $ because, otherwise, it cannot be properly described by the BdG Hamiltonian in~\eqref{eq:H}. Instead, such a strong disorder regime should be studied using the framework of random matrix theory in class D ensemble~\cite{beenakker1997randommatrix,guhr1998randommatrix,brouwer1999distribution,beenakker2015randommatrix,mi2014xshaped,pan2020generic}. Topological MZMs do not exist at all in such highly disordered systems, and we do not study strong disorder in the current work.

\subsection{Numerical method}
To numerically calculate the conductance spectra of the NS junction, we apply the Python scattering matrix transport package KWANT~\cite{groth2014kwant}.
Following the standard procedure, we first discretize the continuum Hamiltonian in Eq.~\eqref{eq:H} into a tight-binding model and then calculate the corresponding S matrix and the conductance as a function of Zeeman field and chemical potential (or gate voltage).

\section{Smooth-potential-induced zero-bias conductance peaks}\label{sec:smooth}
In this section, we focus on the situation where a smooth confinement potential is present near the NS junction inducing a quasi-Majorana bound state in the topologically trivial regime~\cite{vuik2019reproducing, moore2018twoterminal,liu2019conductance}.
We analyze the features of the quasi-Majorana-induced ZBCP and discuss how to distinguish it from the topological Majorana counterpart. 
Furthermore, we perform a self-consistent Thomas-Fermi-Poisson calculation of the electrostatic potential in realistic 3D devices and show that a smooth confinement potential and the induced quasi-Majorana bound state are likely to appear in realistic devices in a smooth potential.

\subsection{Basic properties of smooth-potential-induced ZBCPs}
\begin{figure}[t]
	\centering
	\includegraphics[width=3.4in]{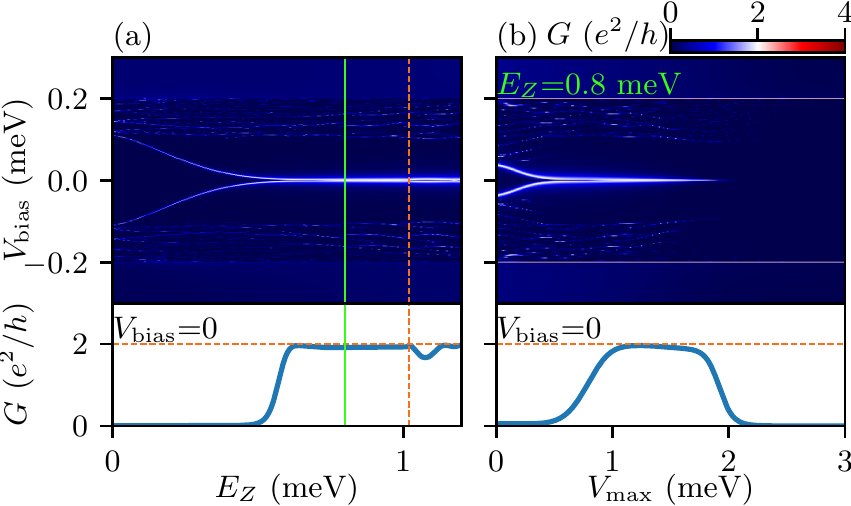}
	\caption{(a) NS conductance as a function of $ E_Z $ and $V_{\text{bias}}$ for a 1D nanowire subject to a Gaussian inhomogeneous potential near the NS junction. The height and width of the potential are $ V_{\text{max}}= $1.2 meV and $ \sigma=0.4~\mu $m, respectively. The red dashed vertical line indicates TQPT. The lower panel shows the corresponding line cut for the conductance at zero bias. (b) Conductance as a function of $ V_{\text{max}} $ and $V_{\text{bias}}$ at a fixed Zeeman field $ E_Z=0.8~$ meV [labeled as the green line in (a)]. The lower panel shows the line cut of the conductance at zero bias as a function of the potential height. The trivial zero-bias conductance shows a stable plateau as a function of both the Zeeman field strength and the smooth potential height, with its maximal value being less than or equal to $2e^2/h$.}
	\label{fig:2}
\end{figure}

We first investigate the basic properties of the smooth-potential-induced ZBCPs.
The smooth potential has a Gaussian-function profile with the spatial width being fixed at $\sigma=0.4~\mu $m, which is roughly 10\% of the wire length.
In Fig.~\ref{fig:2}, we show the calculated conductance as a function of both the Zeeman field strength $E_Z$ and smooth potential height $V_{\text{max}}$.

Figure~\ref{fig:2}(a) shows the tunnel conductance as a function of the bias voltage $V_{\text{bias}}$ and the field strength $E_Z$ at a fixed smooth potential height $ V_{\text{max}}=1.2 $ meV.
The notable feature in Fig.~\ref{fig:2}(a) is a stable conductance plateau of $ 2e^2/h $ within the range of 0.8~meV $< E_Z < $ 1.02~meV.
Such a conductance plateau is induced by the topologically trivial quasi-Majorana because the field strength is less than the critical value $ E_{Zc}= \sqrt{\mu^2 + \gamma^2}= 1.02$~meV for TQPT (labeled as the red dashed line).
Furthermore, as indicated in the lower panel of Fig.~\ref{fig:2}(a), the smooth-potential-induced trivial conductance plateau at zero bias is quite stable as a function of $E_Z$, and is even more stable than the Majorana-induced counterpart for $ E_Z > 1.02 $ meV. For Majorana bound states (MBS), the two Majoranas at the opposite ends of the nanowire overlap and cause oscillation in the ZBCP, while for quasi-Majoranas, both MBS are formed at one wire end overlapping strongly with each other {(see Fig.~\ref{fig:wf} in the Appendix).}

This smooth-potential-induced quantized trivial zero-bias peak is stable not only against the Zeeman field strength but also against the potential height. 
In Fig.~\ref{fig:2}(b), we plot the conductance as a function of the potential height $ V_{\text{max}} $, with the field strength being fixed at $ E_Z=0.8 $~meV less than the critical strength. 
We see that a stable and quantized zero-bias conductance plateau is present between 1~meV $< V_{\text{max}} < 1.8$~meV.
It is due to the quasi-Majorana state induced by the smooth confinement potential.
When the potential height is small ($ V_{\text{max}} < 1$~meV), the potential becomes flat and is no longer confining. Then, the quasi-Majorana gaps out and the zero-bias conductance peak moves to finite bias as shown in Fig.~\ref{fig:2}(b).
On the other hand, when the barrier is large ($ V_{\text{max}} > 1.8$~meV), the confinement potential serves as a strong tunnel barrier. Then, the high barrier reduces the junction transparency suppressing the tunnel conductance to zero.
Therefore we show here that a stable trivial quantized zero-bias conductance plateau at $ 2e^2/h $ is not an exclusive signature of real MZMs; it may also be falsely mimicked by the trivial quasi-Majoranas in the presence of a smooth confinement potential with suitable parameters. Note that at finite temperatures, the ZBCP will have a conductance slightly lower than $ 2e^2/h $~\cite{setiawan2017electron}.

\subsection{Distinguishing between real and quasi-MZMs by correlation measurements}
\begin{figure}[t]
	\centering
	\includegraphics[width=3.4in]{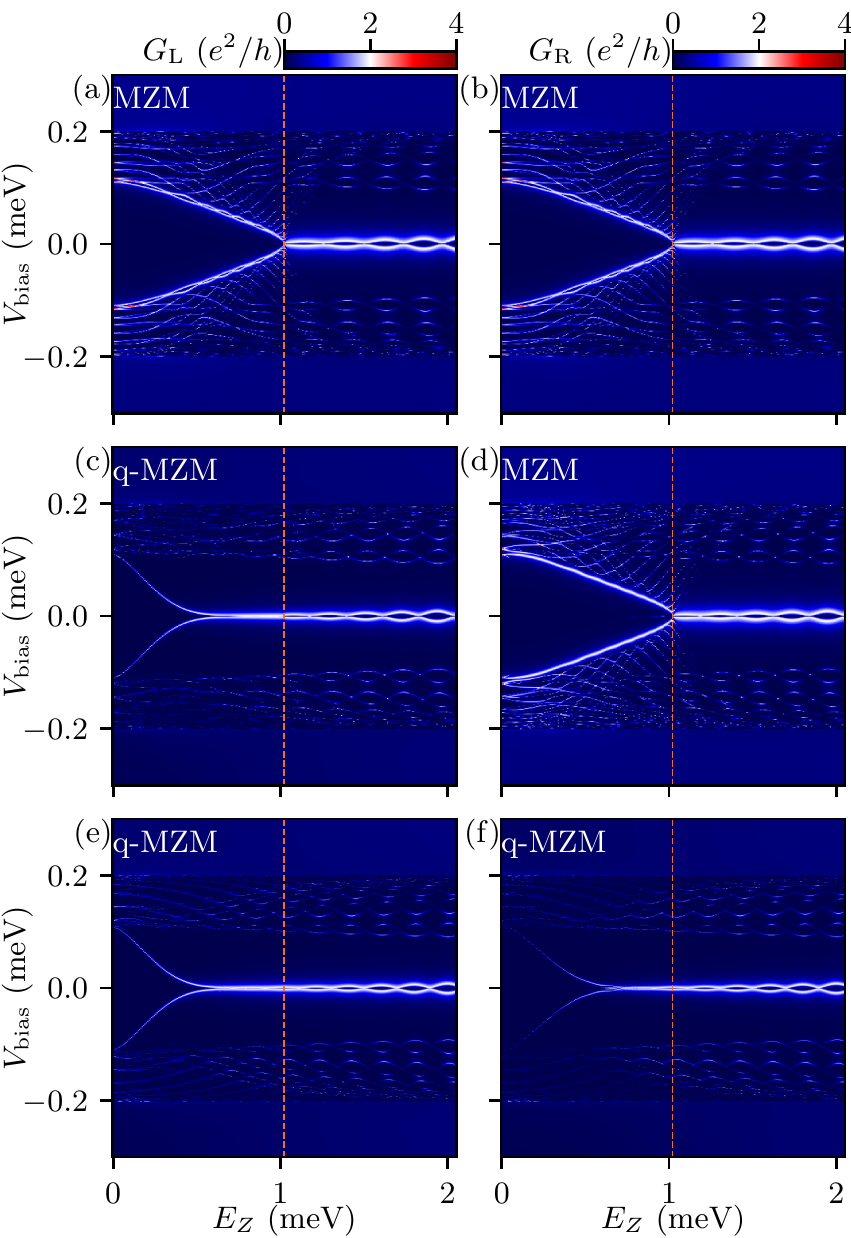}
	\caption{The correlation measurements for three different physical scenarios. Here the conductance is measured as a function of the bias voltage and the strength of the Zeeman field. The left and right columns show the left-end conductance $ G_{\text{L}} $ and right-end conductance $ G_{\text{R}} $.
		(a, b) A pair of MZMs with one MZM being localized at each end. (c, d)  One smooth-potential-induced quasi-Majorana bound state at the left end and one MZM at the right end. The smooth potential at the left end of the nanowire has the parameter of $ (\sigma,V_{\text{max}})=(0.4~\mu\text{m},1.2~\text{meV}) $. (e, f) Two quasi-Majorana bound states, with one at each end of the nanowire. The smooth potentials near the two ends are slightly different. The left potential has the parameters of $ (\sigma,V_{\text{max}})=(0.4~\mu\text{m},1.2~\text{meV}) $, and the right one has the parameters of $ (\sigma,V_{\text{max}})=(0.3~\mu\text{m},1.5~\text{meV}) $.}
	\label{fig:3}
\end{figure}

We now consider a correlation measurement in the three-terminal setup~\cite{pan2021threeterminal,lai2019presence,rosdahl2018andreev} for distinguishing between real topological MZMs and trivial quasi-Majorana bound states.
The correlation measurement can detect the nonlocality of the bound-state wave function. For Majorana bound states, a pair of Majoranas appear at both ends and manifest themselves simultaneously as quantized zero-bias peaks in the {tunneling} spectroscopy. 
By contrast, quasi-Majoranas guarantee the conductance peak at only one wire end because its wave function is localized only at that end.
To illustrate this correlated scheme, we consider three different scenarios: (1) a pristine nanowire, (2) a wire with a smooth potential on the left end, and (3) a wire with smooth potentials on both ends. The calculated left- and right-end conductance results are shown in Fig.~\ref{fig:3}.

The top row of Fig.~\ref{fig:3} shows the topological Majorana conductance of a pristine nanowire, where the left and right conductance spectra are identical. The ZBCPs appear above the same value of Zeeman field strength ($E_Z = E_{Zc}$) at the two ends thus showing perfect correlation, because both ends of the nanowire are occupied by the real MBS above the TQPT (labeled as the red dashed line).

In the middle row of Fig.~\ref{fig:3}, we show the conductance for a nanowire in the presence of a smooth confinement potential localized only at the left wire end and the other end free of any inhomogeneity. In the weak field regime below TQPT ($E_Z < E_{Zc}=1.02~$meV), a stable essentially quantized zero-bias conductance plateau of $ 2e^2/h $ appears only in the left conductance results [see Fig.~\ref{fig:3}(c)], while no ZBCP is observed in the right conductance [see Fig.~\ref{fig:3}(d)]. This is because the localized quasi-Majorana bound state below TQPT appears only at the left end in this scenario of the inhomogeneous potential.
However, when the Zeeman field strength is above the critical value, the whole nanowire becomes topological, and a perfect end-to-end correlation between the left and right conductance spectroscopies appears. This means that whenever a quasi-Majorana situation appears below TQPT, increasing the magnetic field must necessarily lead to the topological MZMs at higher field strength.  There is no experimental report of such an observation in the literature.

In the third scenario, two slightly different smooth confinement potentials are present at the left and right wire ends, inducing quasi-Majoranas at each wire end in the topologically trivial regime below TQPT. As shown in the bottom row of Fig.~\ref{fig:3}, although zero-bias conductance peaks are observed in both the left and right tunnel conductance individually below TQPT, they are not correlated, e.g., the values of $E_Z$ at which ZBCP starts to emerge at the two ends are different. This is because these ZBCPs at $E_Z < E_{Zc}$ are induced by quasi-Majoranas, the properties of which depend sensitively on the details of the smooth confinement potential. For topological MZMs [Fig.~\ref{fig:3}(a)], correlated ZBCPs appear at the same $ E_Z $ for both ends, which is the TQPT point.  
\subsection{Distinguishing between real and quasi-MZMs by stability against Zeeman field}
\begin{figure}[t]
	\centering
	\includegraphics[width=3.4in]{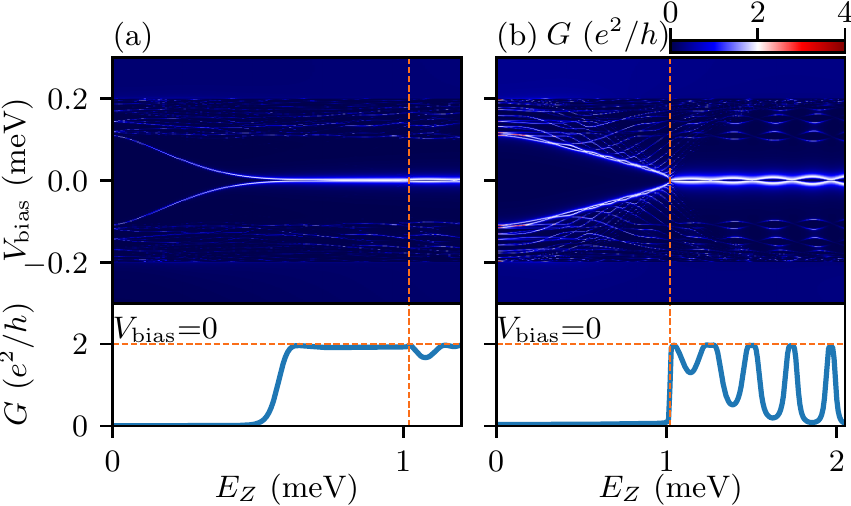}
	\caption{The stability of conductance peaks against the Zeeman field. (a) The conductance for a smooth potential-induced quasi-MZM, which is quite stable against the Zeeman field. Refer to Fig.~\ref{fig:2}(a) for the parameters. (b) The conductance for a real MZM in the pristine nanowire. The zero-bias conductance shows prominent oscillations between 0 to $2e^2/h$ as a function of the Zeeman field strength, owing to the wave function overlap between two Majorana zero modes at the opposite ends of a finite-length nanowire.}
	\label{fig:4}
\end{figure}

We now consider another method for distinguishing between real and quasi-Majorana bound states; which is, by testing the stability of conductance quantization at zero bias against the Zeeman field strength.
In a realistic topological Majorana nanowire, due to the finite-size effect, the wavefunctions of the two localized Majorana bound states at both ends overlap and lead to an oscillation of the zero-bias conductance peak as a function of the increasing Zeeman field strength. 
By contrast, for topologically trivial quasi-Majorana bound states, since its wavefunction is localized at one end of the nanowire, the corresponding zero-bias conductance peak does not show oscillatory behaviors.
In Fig.~\ref{fig:4}, we present the conductance calculations for both trivial quasi-Majoranas (left panels) and topological Majorana bound states (right panels).
The ZBCP induced by the topological MBS in Fig.~\ref{fig:4}(b) oscillates prominently above TQPT with the Zeeman field strength ($E_Z > 1~$meV), and the oscillation amplitude increases with the increasing field strength.
By contrast, as shown in Fig.~\ref{fig:4}(a), the ZBCP induced by the trivial quasi-Majorana bound state is stable against the field strength. 
There are no detectable variations in the quantized conductance plateau in the range of 0.6 meV $< E_Z < 1.02~$meV.
Therefore the observation of an increasing oscillation in ZBCP with the Zeeman field strength can support the presence of Majorana bound states over quasi-Majorana bound states in the hybrid nanowire~\cite{dassarma2012splitting}.
In some situations where smooth potential and q-MZMs appear on both sides of the nanowire, as discussed in the previous subsection, the ZBCP induced by q-MZM would also oscillate [see Figs.~\ref{fig:3}(e) and~\ref{fig:3}(f)].
But the oscillation amplitude is still much smaller than the MBS ZBCPs, because q-MZMs form in the much weaker Zeeman field regime. This also necessitates the presence of q-MZMs on both ends of the wire, which cannot be ruled out but is not a generic situation.

\subsection{Conductance for realistic 3D NS junctions}

In this subsection, we go beyond the 1D minimal model and consider the realistic 3D NS junction, focusing on whether the smooth confinement potential and the induced quasi-Majorana bound state can appear in realistic device geometries with realistic gate voltage values.
Instead of adding the potential profile manually simply as a model potential term $ V(x) $ in Eq.~\eqref{eq:H}, we now calculate the electrostatic potential profile inside the nanowire by solving the 3D Thomas-Fermi-Poisson equation self-consistently, including all the ingredients in a typical NS junction device--- the dielectric layers, the metallic aluminum layer, the normal lead, and the gates. 
As we will show, when the values of the gate voltages are appropriately chosen, a smooth confinement potential can appear near the junction, giving rise to a quasi-Majorana-induced ZBCP in the 3D realistic device at a finite magnetic field. This also provides a justification for the effective 1D model of Eq.~\eqref{eq:H}.

\begin{figure}[t]
	\centering
	\includegraphics[width=3.4in]{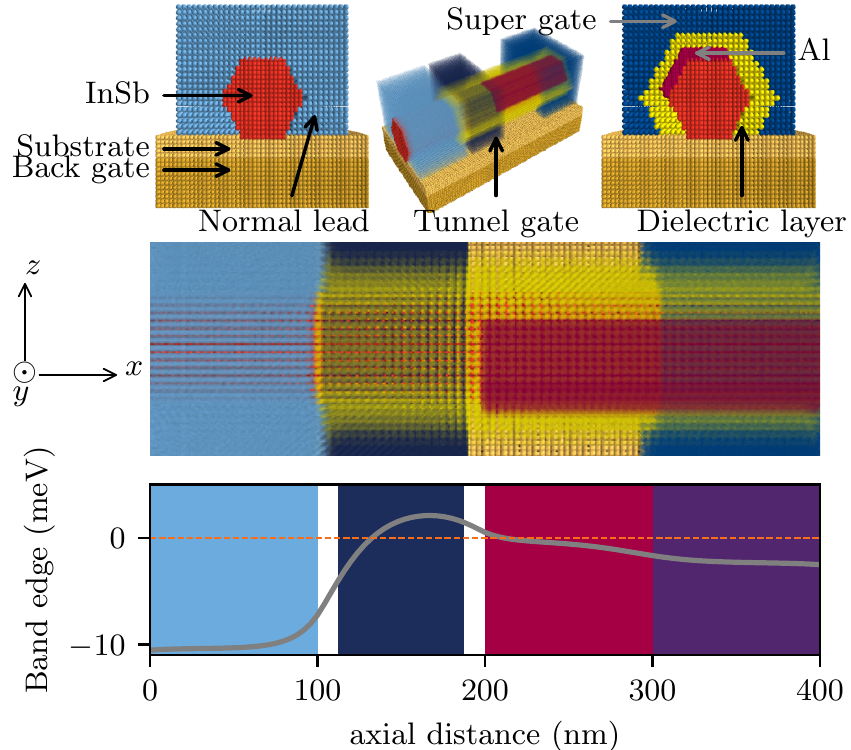}
	\caption{The schematic of the { first 400 nm of the} 3D NS junction. Top panels (left to right): view from the normal lead end, from the side, and from the hybrid nanowire end. Middle panel: view from the top. { The in-plane $ x,z $ directions and out-of-plane $ y $ direction are shown on the left.} Bottom panel: the view of band edge profile {(gray)} focuses  on the first 400 nm of the wire, where the inhomogeneity is prominent.}
	\label{fig:schematic_3d}
\end{figure}

The model system, as shown in Fig.~\ref{fig:schematic_3d}, represents a typical NS junction device in the laboratory, and is used here for the electrostatic potential calculation.
The self-consistent Thomas-Fermi-Poisson equation is~\cite{mikkelsen2018hybridization,antipov2018effects,mikkelsen2018hybridization,woods2018effective,winkler2019unified,liu2020electronic,vuik2016effects}
\begin{align}
\nabla \cdot [ \varepsilon_r(\bold{r}) \nabla \phi(\bold{r}) ] =  -\frac{\rho[ \phi(\bold{r}) ]}{\varepsilon_0},
\label{eq:TFP}
\end{align}
with $\phi$ being the electrostatic potential, and $ \varepsilon_r ( \varepsilon_0)$ the relative (vacuum) permittivity. $\rho$ is the charge density of the mobile electrons in the InSb semiconducting nanowire, which in the Thomas-Fermi approximation is
\begin{align}
&\rho_{\text{e}}(\phi) =  -\frac{e}{3\pi^2 } \left( \frac{2m^* e \phi \theta(\phi)}{\hbar^2}  \right)^{3/2},
\end{align}
where $m^* = 0.015~m_{\text{e}}$ and $\theta(x)$ is the Heaviside step function. We did not include any free holes in the valence band within the Thomas-Fermi approximation because they are irrelevant in the parameter regime of gate voltages in this work, where the Fermi level is always in the conduction band.
The effects of gates are included as Dirichlet boundary conditions inside the gate regions, with the potential values being fixed as the gate voltages.

After obtaining the electrostatic potential inside the semiconducting nanowire, we turn to the quantum mechanical problem by calculating the band edge profile and the conductance in the 3D NS junction. 
The BdG Hamiltonian for conductance calculation is as follows:
\begin{align}
H_{\text{3D}} = & \Bigg[ -\frac{\hbar^2}{2m^*}(\partial^2_x + \partial^2_y + \partial^2_z)  + \alpha_R ( -i \partial_x \sigma_z + i \partial_z \sigma_x )  \nn
&  - e\phi(\bold{r}) \Bigg] \tau_z + E_Z \sigma_x + \Delta(\bold{r}) \tau_x.
\label{eq:ham_3D}
\end{align}
Here $\alpha_R=0.3~\text{eV}\angstrom$ is the strength of the spin-orbit coupling, and the spin-orbit field is pointing along the $ y$-direction. $\phi(\bold{r})$ is the 3D electrostatic potential profile obtained from the Thomas-Fermi-Poisson self-consistent calculation. $\Delta(\bold{r})$ is the induced superconducting gap in the regions covered by the aluminum layer.
{Note that here we have not included all the 3D effects, e.g., the orbital effect of the magnetic field, spatial dependence of the spin-orbit coupling strength, because they would depend rather sensitively on unknown experimental details varying from sample to sample (e.g., the precise magnetic field direction, the precise Al coating on the nanowire, various asymmetric voltages, etc.). Our goal is generic physics, not the simulation of a specific device, which does not vary greatly from sample to sample.}

\begin{figure}[t]
\centering
\includegraphics[width=3.4in]{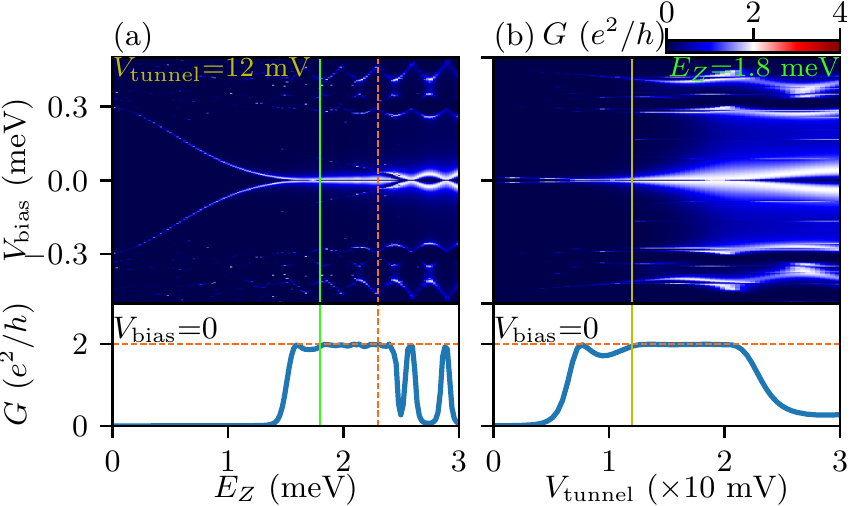}
\caption{The conductance for realistic 3D NS junctions. Here the electrostatic potential profile is numerically calculated using the self-consistent Thomas-Fermi-Poisson method. (a) NS conductance as a function of $ E_Z $ and $V_{\text{bias}}$ with the tunnel gate being fixed at {$V_{\text{tunnel}}=12~$mV}. The lower panel of (a) shows the zero-bias conductance as a function of the Zeeman field strength. The TQPT is labeled by the red dashed vertical line. (b) NS conductance as a function of $ V_{\text{tunnel}} $ and $V_{\text{bias}}$ with the Zeeman field being fixed at {$E_Z=1.8~$meV}. The lower panel of (b) shows the zero-bias conductance as a function of the tunnel gate voltage. Note that the strong similarity between this figure and Fig.~\ref{fig:2} indicates that the physical scenario of smooth potential-induced ZBCP is possible in realistic situations and that simulation based on the 1D effective model is a good approximation for the 3D model in the single-subband limit.}
\label{fig:G_3D}
\end{figure}

Figure~\ref{fig:G_3D}(a) shows the calculated conductance as a function of the field strength and the bias voltage for the 3D NS junction with a particular set of device parameters. The electrostatic potential profile is fixed for Fig.~\ref{fig:G_3D}(a) with the gate voltages being {$V_{\text{Al}}$=17.5 mV (band offset between InSb and Al), $V_{\text{supergate}}$=26 mV, $V_{\text{tunnel}}$=12 mV, $V_{\rm{lead}}$=27 mV, and $V_{\text{backgate}}$=0.}
We further visualize the 3D electrostatic potential by plotting the 1D band edge profile in the vicinity of the junction along the wire axis in the bottom panel of Fig.~\ref{fig:schematic_3d}.
The band edge profile is for the lowest subband in the nanowire, that is, the hybrid nanowire is in the single-subband limit.
Note that a smooth confinement potential is naturally present between 100~nm $< x < $ 300~nm, owing to the combined effect of the tunnel gate, the InSb-Al band offset, and the supergate above the nanowire. The smooth potential here is present by virtue of the electrostatics of the 3D NS junction without having to be put in manually as a model potential as in the effective 1D model of Eq.~\eqref{eq:H}.
The nanowire is {$ 3~\mu$m} in total length, and the potential value reaches an asymptotic value for $x>400~$nm; that is, we take $\phi(x>400~\text{nm}, y, z) = \phi(x =400~\text{nm}, y, z) $. So, the smooth potential extends over roughly 10\% of the wire near the tunnel junction end.
In Fig.~\ref{fig:G_3D}(a), a quasi-Majorana-induced nearly-quantized zero-bias conductance peak appears in the trivial regime (to the left of the red dashed line), due to the presence of the smooth confinement potential.
The stable quantized zero-bias conductance peak becomes oscillatory after the nanowire enter the topological phase at a stronger Zeeman field (to the right of the red dashed line).
We then fix the strength of the Zeeman field at {$E_Z=1.8~$meV}, which is below the TQPT, such that the ZBCP is now induced by the quasi-Majorana bound state, and sweep the voltage of the tunnel gate.
The resulting conductance profile as a function of the tunnel gate voltage is shown in Fig.~\ref{fig:G_3D}(b).
Note that the quasi-Majorana-induced ZBCP is stable and nearly-quantized only in the regime { $10~\text{mV}<V_{\text{tunnel}}<20~$mV.}
Outside this range of tunnel gate voltages, the confinement potential at the junction is either too sharp or too flat to induce a quasi-Majorana bound state.
Figure~\ref{fig:G_3D}, therefore, shows that a smooth confinement potential is indeed possible in a realistic 3D NS junction device with appropriate gate voltage values and that such potential can induce a quasi-Majorana bound state with a stable almost-quantized ZBCP in the {tunneling} spectroscopy. Thus, the observation of a `stable' and `quantized' ZBCP is not sufficient to establish the existence of topological MBS in the system.

Another important finding here is that the conductance profiles obtained from 3D simulations (Fig.~\ref{fig:G_3D}) closely resemble those from the effective 1D model (Fig.~\ref{fig:2}), especially the conductance line cuts at zero bias. 
This indicates that the 1D minimal model is a good approximation of the 3D realistic nanowire model when the nanowire is in the single-subband limit.
This justifies all the calculations using the 1D minimal model in this and other previous works. Given the very high computational cost of the 3D simulations and the relative simplicity of the 1D calculations, it makes sense to base conductance calculations on the 1D model.

\section{Dip-potential-induced zero-bias conductance peaks}\label{sec:dip}
\begin{figure}[t]
	\centering
	\includegraphics[width=3.4in]{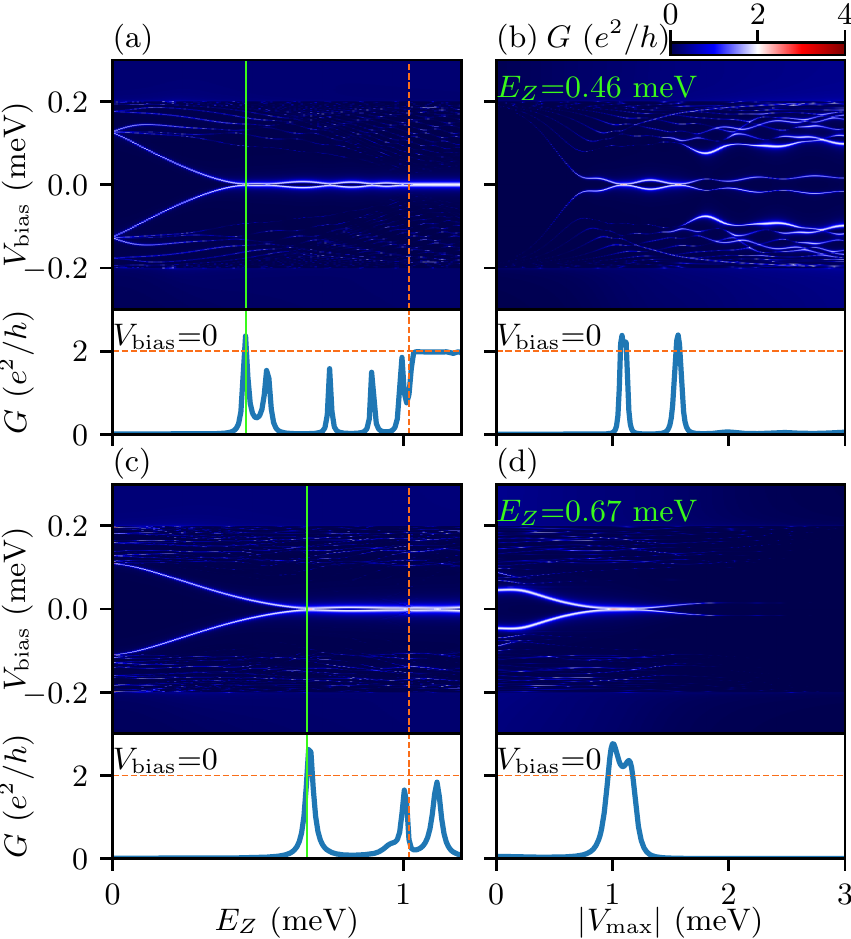}
	\caption{
		The conductance for dip-like potential-induced Andreev bound states.
		One generic feature of the conductance in this physical scenario is that the ZBCP as a function of the system parameter shows a spike-like conductance peak which can be greater than $ 2e^2/h $.
		(a) The conductance as a function of $ E_Z $ and $V_{\text{bias}}$ for a nanowire which has a negative chemical potential and is subject to a dip potential near the NS junction. Here it takes the form of Gaussian potential with the parameters of $ (\sigma,V_{\text{max}})=(0.5~\mu\text{m},-1.1~\text{meV}) $, and the chemical potential $ \mu=-1 $ meV. The lower panel is the line cut for the conductance at zero bias. (b) The conductance as a function of depth $ V_{\text{max}} $ and $V_{\text{bias}}$ for a nanowire with a negative chemical potential at $ E_Z=0.46 $ meV [labeled in the green dashed line in (a)]. (c) Similar to (a), but now the dip-like potential is in the form of the sinusoidal potential, with the parameter being $ (\sigma,V_{\text{max}},\sigma^{\prime},V_{\text{max}}^{\prime})=(0.4~\mu\text{m},1~\text{meV},0.2~\mu\text{m},1~\text{meV})$. The chemical potential of the nanowire is $ \mu=1 $ meV. (d) Similar to (b) for the sinusoidal potential. The strength of the Zeeman field is fixed at $E_Z=0.67 $ meV [labeled in the green dashed line in (c)].}
	\label{fig:7}
\end{figure}

In this section, we consider, within the 1D effective model of Eqs.~\eqref{eq:H}-\eqref{eq:sinusoidal}, a quantum-dot-like potential near the NS junction, i.e., a dip in the potential, rather than the smooth barrier-like potential considered above.
We focus on the peak values and the conductance robustness of the induced ZBCPs.
In particular, the quantum dot considered here is either a Gaussian-function potential with a negative amplitude (a dip in the potential profile) or a sinusoidal-function potential (armchair-like potential profile), as shown in Figs.~\ref{fig:1}(b) and~\ref{fig:1}(c).
Under these conditions, low-energy Andreev bound states appear inside the potential inhomogeneity near the end of the hybrid nanowire at a finite strength of the Zeeman field. These trivial Andreev bound states can induce (near-) zero-bias conductance peaks, mimicking topological Majorana bound states. However, a major difference between Andreev and Majorana bound states for the dip-potential is that the zero-bias conductance profiles of trivial peaks are sharp spike-like in the parameter space, e.g., in $E_Z$ or $\mu$, and the peak values may exceed $2e^2/h$ in extremely narrow ranges of parameters, while those of topological peaks are plateau-like with peak values never more than $2e^2/h$.

Figure~\ref{fig:7}(a) shows the calculated ABS-induced trivial conductance as a function of bias voltage and Zeeman energy. The quantum-dot potential is dip-like with depth $V_{\text{max}} = -1.1~$meV, linewidth $\sigma=0.5~\mu$m, and the bulk chemical potential $\mu=-1$~meV. A zero-bias conductance peak forms in the trivial regime $0.45~\text{meV} \lesssim E_{Zc} \lesssim 1~$meV, owing to the quantum-dot-induced Andreev bound states near the NS junction. The zero-bias conductance peak oscillates as the strength of the Zeeman field increases, somewhat resembling the Majorana-induced peak oscillation [Fig.~\ref{fig:4}(b)]. However, at $E_Z \approx 0.46~$meV, the value of the zero-bias conductance peak exceeds $2e^2/h$, which is a unique feature that distinguishes these dot-induced trivial ABSs from topological MBSs. 
We also calculate the conductance as a function of the dot potential amplitude $V_{\text{max}}$ at a fixed Zeeman field strength, which qualitatively models the tunnel-gate-dependence of conductance in realistic junctions. As shown in Fig.~\ref{fig:7}(b), an oscillatory conductance peak near zero bias appears in the range of $-1.8~\text{meV} < V_{\text{max}} < -0.8~$meV, and more importantly, its peak value is greater than $2e^2/h$ at $V_{\text{max}} \approx -1~$meV and $-1.5~$meV.

Figures~\ref{fig:7}(c) and~\ref{fig:7}(d) show the calculated conductance in the presence of a sinusoidal-function-like potential [see Eq.~\eqref{eq:sinusoidal}] near the junction. Similar to the dip-like-potential scenario, a topologically trivial ZBCP forms at finite strength of Zeeman field before TQPT ($E_Z < E_{Zc}$). Furthermore, the zero-bias peak profile at $E_Z \approx 0.67~$meV and $V_D \approx 1~$meV is spike-like instead of plateau-like, and the peak value exceeds $2e^2/h$.

One way to understand the features of dip-potential-induced `spiky' ZBCPs is to decompose the low-energy Andreev bound states into a pair of overlapping Majorana wave functions forming the q-MZMs {[see Fig.~\ref{fig:wf}(d) in the Appendix]}.
The presence of the dip potential enhances the coupling between the two Majoranas wave functions, and thus the energy of the Andreev bound state crosses the zero energy instead of sticking. 
Therefore we see only spike-like ZBCP rather than a conductance plateau in this scenario.
Moreover, owing to the dip potential, both Majoranas couple effectively with the external normal lead, which causes the tunneling peak value to be greater than $2e^2/h$. In fact, our simulations show (see the animations presented in the Supplemental Material~\cite{movie}) that the resulting zero-bias conductance quickly rises to $ 4e^2/h $, corresponding to the combined conductance of two Majoranas, and then quickly becomes zero as the ABS is no longer a zero-energy state. An extremely fine-tuned dip-potential may give rise to a zero-bias conductance above $ 2e^2/h $, but it is unlikely to be experimentally observable as a ZBCP because of the strongly spiky nature of the conductance peak.  This is in sharp contrast to the barrier-like potential where a conductance plateau at $ 2e^2/h $ may stick to zero bias for finite parameter regimes in both the Zeeman field and chemical potential.

\section{Disorder-induced zero-bias conductance peaks}\label{sec:disorder}

\begin{figure}[t]
	\centering
	\includegraphics[width=3.4in]{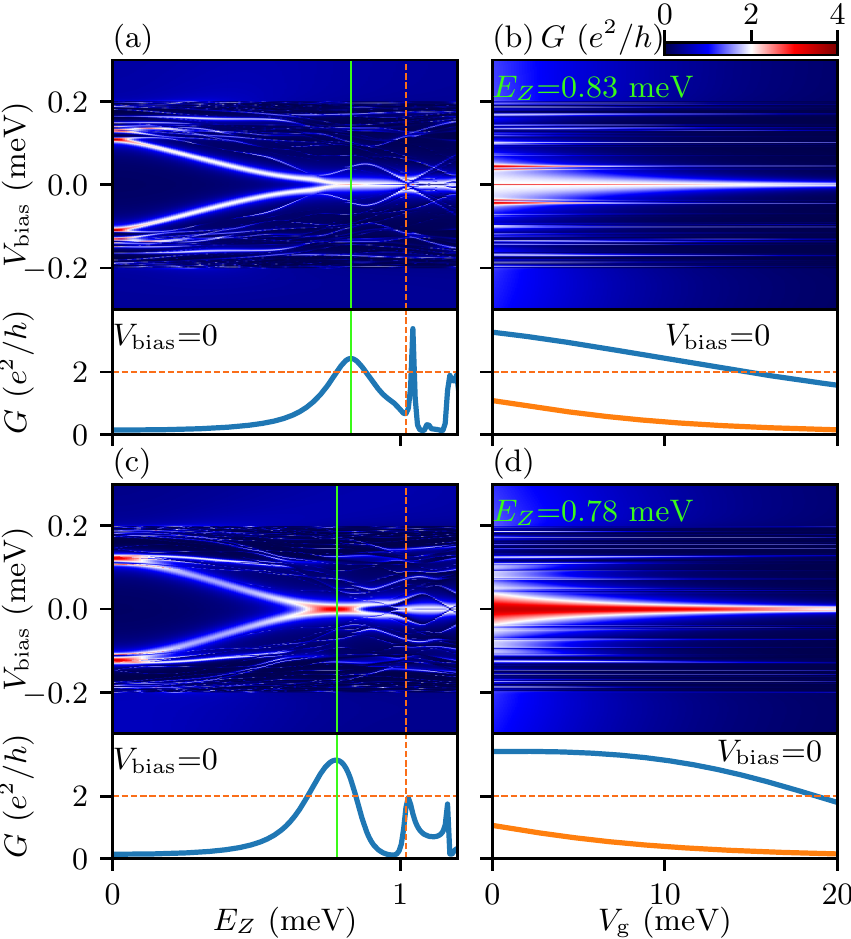}
	\caption{Conductance for a Majorana nanowire in the presence of a random disorder potential in the bulk.
		(a) Conductance as a function of $ E_Z $ and $V_{\text{bias}}$. The disorder potential can induce ZBCP when the Zeeman field is strong enough, and the zero-bias conductance as a function of the Zeeman field strength shows a spike-like profile with the peak value greater than $ 2e^2/h $. 
		(b) The peak conductance at zero bias (blue solid line) and normal conductance above SC gap ({orange} solid line) as a function of $ V_g $ and $V_{\text{bias}}$ at fixed $ E_Z=0.83 $ meV [labeled in the green line in (a)]. The conductance increases monotonically with the decreasing potential height. (c, d) Conductance as a function of $ E_Z $ and $ V_{\text{bias}} $ under another set of disorder configuration, which also manifests the same conductance peaks.}
	\label{fig:8}
\end{figure}

We now consider the scenario in which the potential inhomogeneity is in the form of a nondeterministic random potential, i.e., disorder.
In realistic devices, such a potential can originate from the growth imperfection at the superconductor-semiconductor interface or from unintentional (and hence unknown) impurities in the wire or the superconductor or the substrate.
Our focus is on whether it is possible to find a disorder-induced apparent trivial quantized conductance plateau whose peak value is greater than $ 2e^2/h $. 
Figure~\ref{fig:8} shows the calculated tunneling conductance in the presence of disorder with two distinct random configurations. 
{We choose the amplitude of the disorder fluctuations to be $\sigma_{\mu}=1$ meV, which is the same order of the chemical potential $\mu=1$ meV.}
The first row of Fig.~\ref{fig:8} uses one specific random configuration. We notice that the gap closes at $ E_Z\sim0.8 $ meV below the TQPT (the red dashed line), and an almost-quantized conductance plateau forms. This plateau of conductance is not as robust as the inhomogeneous-potential-induced plateau, as shown in Fig.~\ref{fig:4}(a). However, if we rescale the $ E_Z $ axis and focus on only the nearby region of the plateau, we can still manually create a `robust' plateau, although this trivial plateau is a deceptive visual effect because of fine-tuning the parameter range. Such a trivial disorder-induced ZBCP `plateau' may misleadingly masquerade as evidence supporting the presence of topological MBS since experimentally the TQPT location is not known,~\cite{pan2019curvature} and \emph{a priori} there is no way to know whether one is below or above the TQPT based just on NS tunneling measurements.

However, this artificial plateau, which to some extent is robust against Zeeman field strength, is not robust against the tunnel barrier. In Fig.~\ref{fig:8}(b), we show the conductance dependence on the tunnel barrier height at a fixed field strength $ E_Z=0.78$ meV [corresponding to the green solid line in Fig.~\ref{fig:8}(a)]. The conductance decreases from $ 3e^2/h $ to $ 1.5e^2/h $ monotonically as $V_g$ increases. This instability indicates that the trivial ZBCP does not have a truly stable quantized conductance with a value higher than $ 2e^2/h $ against both the Zeeman field and the tunnel barrier. The important point to note is that the disorder-induced trivial ZBCP could produce a somewhat-stable plateau with conductance above $ 2e^2/h $.

In the second row of Fig.~\ref{fig:8}, we present the conductance results for another random disorder configuration. A trivial ZBCP with a peak value well above $2e^2/h$ is observed at $E_Z \sim 0.7~$meV, similar to the conductance profile in Fig.~\ref{fig:8}(a). Interestingly, when we fix the field strength at $E_Z \sim 0.7~$meV and vary the tunnel barrier height, the ZBCP shows a robust plateau between 0 meV $< V_g < 5~$meV whereas the above gap conductance changes by more than 50\%, before it drops at larger values of $V_g$ [see Fig.~\ref{fig:8}(d)]. 
Note that the ZBCPs in Fig.~\ref{fig:8} for both random configurations are obtained by careful fine-tunings { and postselection}. 

\begin{figure}[t]
	\centering
	\includegraphics[width=3.4in]{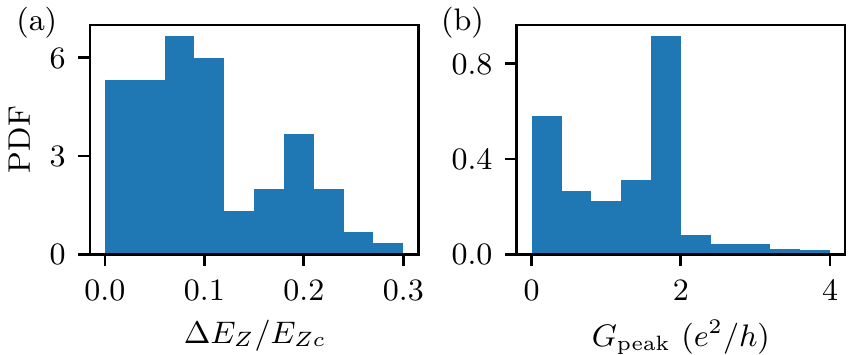}
	\caption{(a) The statistics of the occurrence of ZBCP as a function of $\Delta E_Z / E_{Zc}$.  The $ y $ axis is the probability density function (PDF). (b) The statistics of conductance given the trivial ZBCP is already found. The parameters are the same as Fig.~\ref{fig:8}, and ensemble size is 100.}
	\label{fig:stat}
\end{figure}

{Furthermore, we consider the properties of disorder-induced ZBCPs from a statistical perspective. Our goal is to show the occurrence probability of the peak stability and the peak value. In particular, we consider an ensemble of disorder potential with fluctuation amplitude $ \sigma_\mu=1 $ meV, and randomly sample 100 different disorder configurations. In each specific conductance spectroscopy, we define two quantities for the conductance peak in the topologically trivial phase ($E_Z < E_{Zc}$): (1) the dimensionless ratio of the energy interval of Zeeman field that ZBCP can appear to the total energy interval of trivial phase, denoted by $\Delta E_Z / E_{Zc}$, which characterizes the stability of a disorder-induced ZBCP, (e.g., a special case without trivial ZBCP corresponds to $\Delta E_Z / E_{Zc}=0$), and (2) the conductance of the peak of the ZBCP given a ZBCP is already found in the tunneling conductance. 
The results are shown in Fig.~\ref{fig:stat}, where Fig.~\ref{fig:stat}(a) shows that most disorder-induced peaks have a ratio less than 0.2, which is consistent with the conductance spectroscopies shown in Fig.~\ref{fig:8}. 
It alludes that disorder-induced ZBCPs are mostly spike-like, and thus have a relatively small ratio.
Figure~\ref{fig:stat}(b) shows that the peak values disorder-induced ZBCPs are mostly below $2e^2/h$, but there are rare occasions where the peak values go above $2e^2/h$. The percentage of ZBCPs going above $2e^2/h$ is roughly 8\% when the strength of disorder is $ \sigma_\mu =1 $ meV.}

These examples demonstrate that disorder could generically introduce trivial ZBCPs at conductance values higher than $ 2e^2/h $, which, with some { postselection and} fine-tuning, may reflect some limited apparent `robustness' in both the applied magnetic field and applied gate voltage depending on the details of the disorder. Of course, disorder can also induce trivial ZBCPs with conductance values at or below $ 2e^2/h $, as has recently been studied in the literature~\cite{pan2020physical,pan2020generic}.

\section{Discussion}\label{sec:discussion}
Strictly speaking, the confirmation of topological superconductivity and MZMs should be based on various experimental measurements probing both the ends and the bulk of the wire simultaneously. For example, the Majorana-induced robust quantized ZBCP should be accompanied by the topological gap closing and reopening in the vicinity of TQPT, and the ZBCP oscillations with an increasing Zeeman field strength. However, in practice, the topological bulk gap features are difficult to observe using the local probe in NS {tunneling} spectroscopies, and the strength of the Zeeman field is constrained by the critical field of the parent superconducting layer above which all bulk superconductivity vanishes, thus destroying the topological superconductivity in the process. The fundamental problem underlying local tunneling spectroscopy is its inability to confirm any signature of nonlocality, which is the hallmark of topological Majorana modes.

In most of the recent experimental measurements, we only see the formation of a ZBCP just before the closing of the parent superconducting gap.
This apparent gap closing feature in the NS tunneling spectrum may, however, have nothing to do with a bulk gap closing, but only a manifestation of Andreev bound states coming together leading to a trivial ZBCP~\cite{liu2017andreev}.
Thus, the question that motivates this work is: What can we learn if we have merely the local tunneling measurements on the Majorana nanowires, and can we infer the underlying physical mechanisms for the ZBCP based on their peak profiles?

With extensive numerical simulations, we find that true- or quasi-Majoranas do (not) induce ZBCPs of a peak value equal to or smaller (greater) than $2e^2/h$ in the zero-temperature limit. 
By contrast, for those ZBCPs due to dip-potential-induced ABSs and disorder-induced bound states, the peak value can exceed $2e^2/h$ generically, although such dip-induced trivial ZBCPs have very sharp spiky structures as a function of system parameters (e.g., Zeeman field, gate potentials) and are therefore observed only when tunnel/temperature/dissipation broadening effect is large.

We further note that the disorder scenario has much richer physics than realized before, because, through careful { postselection and} fine-tunings, a particular random disorder configuration may induce a ZBCP (above or below or close to $ 2e^2/h $) which shows robustness against either the field strength or the tunnel barrier height.
So any NS tunneling observation by itself, e.g., a nearly quantized ZBCP at or around $ 2e^2/h $, or a robust peak against one system parameter, is not sufficient to conclude topological Majoranas or even trivial quasi-Majoranas in superconductor-semiconductor nanowires.
Only with a combination of conductance quantization at $2e^2/h$ at zero temperature, robustness against both Zeeman field strength and gate voltages, the observation of Majorana oscillations with increasing field strength, closing/reopening of the bulk gap just as the ZBCP develops, and correlation measurements on both wire ends can give convincing evidence to Majorana zero modes. One needs to be particularly mindful of the fact that disorder-induced trivial ZBCPs may have observable features above, at, or below $ 2e^2/h $, perhaps even manifesting some apparent (but misleading) limited robustness in the Zeeman field and chemical potential.

We also emphasize that all the calculations in this work are based on the single-subband assumption. The number of subbands in the devices in the laboratory is not known \emph{a priori}. When multiple subbands come below the Fermi level of the hybrid nanowire due to gating, the ZBCP may exceed $2e^2/h$ in principle~\cite{wimmer2011quantum}. However, even in the multi-subband model, a conductance plateau is observed only at half-integer multiples of $4 e^2/h$, and a plateau of conductance slightly above $2e^2/h$ is unlikely for MZM.

Before concluding, we now provide a brief discussion of the existing experimental results on Majorana nanowires in the context of our theoretical findings.  Early Majorana experiments during 2012 to 2016 were all plagued by very strong disorder in the systems leading to soft gaps indicating the presence of considerable disorder-induced subgap fermionic states, and although weak ZBCPs ($ \sim 0.1-0.2 e^2/h $) often manifested in the tunnel conductance spectra, it is manifestly clear that these experimental observations are not conclusive at all for the existence of topological Majorana modes. The first experiment involving a hard gap was by Deng \emph{et al}.~\cite{deng2016majorana} which saw the emergence of ZBCPs (but with peak height $ \ll 2e^2/h $) arising from the merging of Andreev bound states.  A detailed theoretical analysis established these observations as most likely the manifestation of trivial ZBCPs with ABSs becoming almost zero-energy states mimicking as MZMs~\cite{liu2017andreev}.  Two influential later papers, by Zhang \emph{et al}.~\cite{zhang2021large,zhang2018quantizeda} and Nichele \emph{et al}.~\cite{nichele2017scaling}, reported the observation of ZBCPs with peak heights at $ 2e^2/h $ in nanowires with hard gaps, creating considerable excitement that perhaps the topological MZMs have been seen.  It was soon realized, however, that the Nichele experiment~\cite{nichele2017scaling} saw ZBCPs with peak heights higher than $ 2e^2/h $ even at finite temperatures (and the $ T=0 $ peak height would be still higher), indicating that the observation is inconsistent with topological MZMs~\cite{setiawan2017electron}. Very recent experimental work~\cite{zhang2021large,zhang2018quantizeda} now indicates that the same is true for the Zhang experiment~\cite{zhang2021large,zhang2018quantizeda} also, with the ZBCP height actually being $ 2.2 e^2/h $ already at $ T=25 $ mK.  In addition, the new data and analysis~\cite{zhang2021large,zhang2018quantizeda} indicate that the original experimental conductance plateau reported in Ref.~\onlinecite{zhang2021large,zhang2018quantizeda} is an artifact and careful consideration of charge jumps in the system eliminates the ZBCP stability.  Therefore, both of these experiments, Nichele~\cite{nichele2017scaling} and Zhang~\cite{zhang2021large,zhang2018quantizeda}, report unstable ZBCPs with conductance values somewhat above $ 2e^2/h $.  Our work demonstrates that these observations are inconsistent with either the topological MZM or the trivial q-MZM interpretation.  Thus, the only possible conclusion, which has also been reached recently in other theoretical publications,~\cite{pan2020physical,pan2020generic,pan2021threeterminal} is that even the best experimental samples of today still have considerable disorder in them, and in all likelihood, ZBCPs being observed experimentally arise from disorder and are trivial.  Given that semiconductor nanowire systems are undoubtedly the cleanest platforms for Majorana experiments, it is reasonable to conclude that all ZBCP-implied conclusions about the observation of Majorana zero modes are actually observing disorder-induced trivial zero-bias peaks {in superconductor-semiconductor heterostructures}. This is in fact an encouraging scenario for the future of Majorana experiments--- all one needs to do is to remove disorder and produce cleaner samples in order to obtain topological Majorana zero modes.

\section{Conclusion}\label{sec:conclusion}
To conclude, we have systematically investigated ZBCPs in the NS {tunneling} spectroscopy arising from different physical mechanisms, focusing on the conductance peak values and their robustness against system parameters. We find that for true topological Majoranas and smooth-potential-induced quasi-Majoranas, the induced ZBCP does not exceed $2e^2/h$.
In a realistic finite-length nanowire, the Majorana-induced ZBCP oscillates with an increasing Zeeman field strength due to the wave function overlap between the Majoranas at the opposite ends of the wire, while the quasi-Majorana-induced ZBCP does not oscillate since its wave function is localized at the end of the inhomogeneity.
Both MZMs and q-MZMs show a conductance plateau as a function of the tunnel barrier transparency.

Beyond the simple single-subband calculation, we use the self-consistent Thomas-Fermi-Poisson method and show that a smooth confinement potential is indeed likely to form near the NS junction due to the combined effects of gates and superconductor-semiconductor band offset.
It also indicates that the realistic 3D model is qualitatively equivalent to the widely used effective 1D minimal model in the single-subband limit.

If there is a dip or well inside the potential inhomogeneity, the ZBCP can exceed $2e^2/h$, and the profile of the ZBCP then becomes spike-like as a function of either Zeeman energy or tunnel barrier height. We ascribe the spike-like conductance to the enhanced inter-Majorana coupling in the dip-like potential inhomogeneity. Such a spiky conductance rises and falls too quickly as a function of parameters (we provide animations in the Supplemental Material~\cite{movie}) to be experimentally relevant in our view. It is, of course, possible that very careful fine-tuning plus effects of temperature and dissipation may lead to just the correct experimentally observed conductance slightly above $ 2e^2/h $~\cite{nichele2017scaling,zhang2021large,zhang2018quantizeda} arising from a quantum dot-induced dip potential, but such fine-tuning is more likely to lead to a ZBCP with conductance below $ 2e^2/h $~\cite{deng2016majorana,liu2017andreev}. In any case, the dip potential would not lead to any robustness in the ZBCP, and cannot be construed to belong to the q-MZM category.

Finally, we present conductance calculations for the random disorder potential which also induces trivial ZBCP, with its peak value which may exceed $2e^2/h$, and having certain limited robustness against either Zeeman field or tunnel gate, if any, through a careful fine-tuning procedure. Such fine-tuned somewhat robust ZBCPs with conductance values above $ 2e^2/h $ have been observed in recent experiments,~\cite{nichele2017scaling,zhang2021large,zhang2018quantizeda}, and we believe that these are all disorder-induced trivial ZBCPs.

In summary, our message to the experimentalists is that the presence of a conductance plateau approaching quantized $ 2e^2/h $ from below without much fine-tunings may indicate the existence of real or quasi-Majorana zero modes. 
By contrast, a conductance peak above $ 2e^2/h $, whether the profile is spike-like or plateau-like, would be from disorder- or dip-induced trivial Andreev bound states. Producing cleaner samples with less disorder should be the highest priority for progress in the field.

The definitive evidence for the existence of topological MZMs in the tunneling spectroscopy must minimally satisfy the following five criteria: (1) generic observation of $ 2e^2/h $ (or slightly below) ZBCP without extensive fine-tuning; (2) stability of the ZBCP in the Zeeman field, tunnel barrier potential, and chemical potential; (3) ZBCP manifesting some oscillatory behavior with increasing Zeeman field; (4) observation of end-to-end tunneling nonlocal correlations in the ZBCP; (5) and observation of a bulk gap closing and reopening concomitant with the appearance of the ZBCP.

Finally, we comment on the nonexistence of any mesoscopic conductance fluctuations in Majorana nanowires, particularly in view of our conclusion that even the best available samples are currently disorder-dominated.  It is worthwhile to mention that our exact phase-coherent tunnel conductance calculations in the presence of random disorder in the nanowire do not manifest any mesoscopic fluctuations in the system. In particular, the theoretical tunnel conductance does not fluctuate randomly at all [let alone by $ O(e^2/h) $] as a function of bias voltage, chemical potential, and Zeeman field; instead, varying smoothly according to the spectral evolution of the BdG equation with parameter variation. In fact, no mesoscopic fluctuations are expected here because of the existence of the superconducting gap and the nature of tunneling transport as would have happened in a similar disordered normal system for coherent diffusive transport. This lack of any disorder-induced mesoscopic fluctuations in theory is consistent with the experimental findings in nanowire tunneling transport, where mesoscopic fluctuations have never been reported despite all the early experimental samples having substantial disorder, as reflected in their having very soft gaps indicating the existence of considerable sample disorder.  In the superconducting systems, where Majorana experiments are carried out, one expects only disorder-induced nonuniversal subgap bound states fluctuating from one disorder configuration to another, as is also found in the theoretical calculations with disorder. The existence of different sets of subgap bound states in different disorder configurations is the analog of conductance fluctuations in the current systems.

This work is supported by the Laboratory for Physical Sciences (LPS), by a subsidy for Top Consortia for Knowledge and Innovation (TKl toeslag) by the Dutch Ministry of Economic, and by the Netherlands Organisation for Scientific Research (NWO/OCW) through VIDI Grant No. 680-47-537 Affairs. We acknowledge the University of Maryland High Performance Computing Cluster (HPCC). 

\bibliography{Paper_inhomogeneity}

\appendix

\section{Wave function}
\begin{figure}[ht]
	\centering
	\includegraphics[width=3.4in]{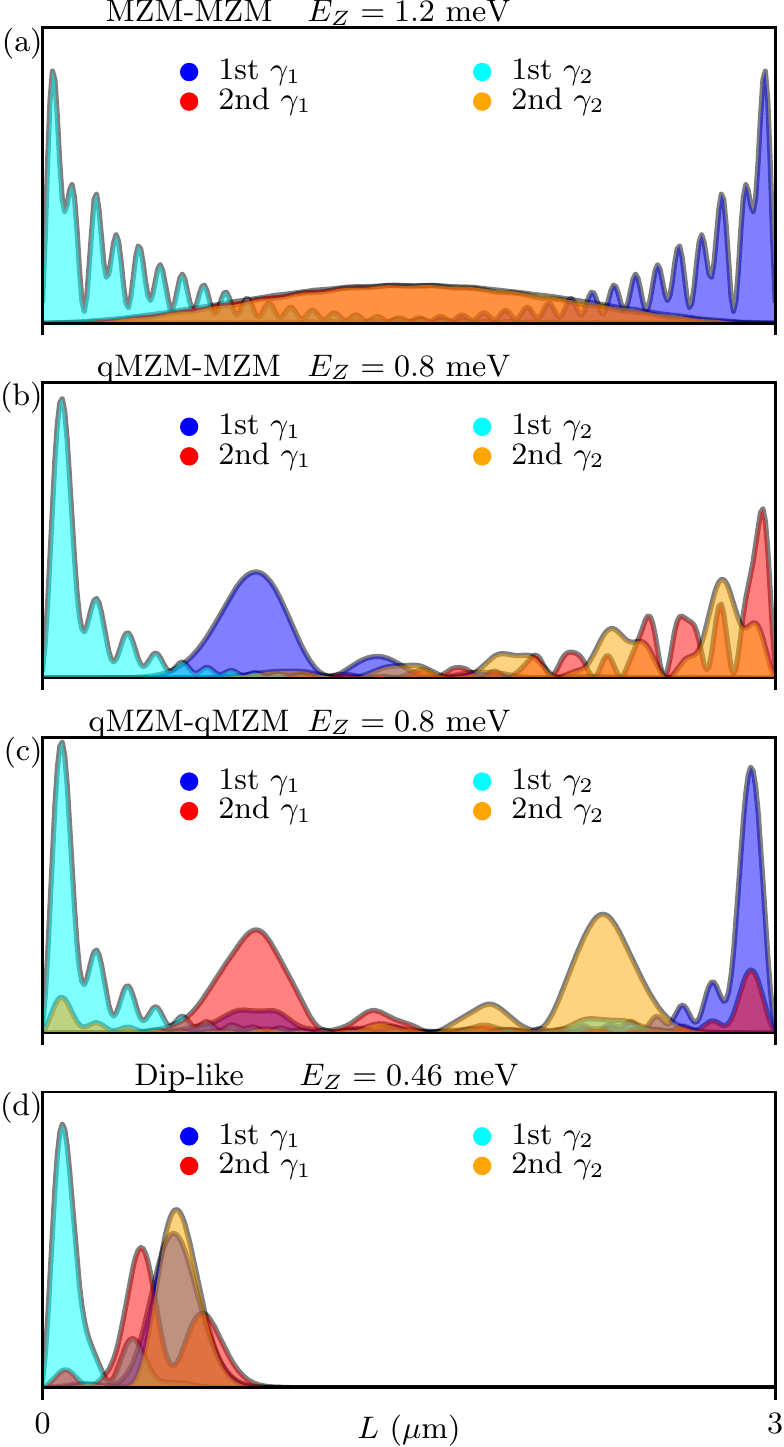}
	\caption{(a) The wave function of two MZMs on both ends at $ E_Z $=1.2 meV; (b) The wave function of the q-MZM on the left and MZM on the right at $ E_Z $=0.8 meV; (c) The wave function of two q-MZM on both ends at $ E_Z $=0.8 meV. (d) The wave function of a dip-like potential corresponding to Fig.~\ref{fig:7} at $ E_Z $=0.46 meV. }
	\label{fig:wf}
\end{figure}

Figure~\ref{fig:wf} shows spatial wave functions for three scenarios corresponding to Fig.~\ref{fig:3}, and one case of dip-like potential corresponding to Fig.~\ref{fig:7}.
\end{document}